\documentstyle[]{mn}

\begin{document}
 
\title[Galactic axes orientations in galaxy clusters]{Investigation of
galactic alignment in LSC galaxy clusters}
 
\author[W. God\l owski \& M. Ostrowski]{W. God\l owski \& M. Ostrowski \\
Obserwatorium Astronomiczne, Uniwersytet Jagiello\'{n}ski, ul. Orla 171,
30-244 Krak\'{o}w, Poland \\
(E-mail: godlows{@}oa.uj.edu.pl \& mio{@}oa.uj.edu.pl)}
 
\date{}
 
\maketitle
 
\begin{abstract}
We investigate  the  galactic  axes  orientations  within  18  selected
clusters,  sub-structures  of  the  Local  Supercluster. For every
cluster we  map the parameter $\Delta_{11}$ (Flin \& God{\l}owski 1986)
describing  the galactic axes alignment with respect  to  a chosen
cluster pole, divided by its formal error $\sigma(\Delta_{11})$ ($s
\equiv \Delta_{11} / \sigma(\Delta_{11})$). The cluster pole
co-ordinates change  along the entire  celestial sphere.  The  resulting
maps  are  analyzed  for correlations of its maxima  with directions
from the cluster centre to 1.) the derived `physical' cluster poles, 2.)
the Local Supercluster centre, 3.) the Virgo A centre and 4.) the Earth,
i.e. along the line of  sight ($\equiv$  LOS).  The strong  maxima --
with one exception -- exist only for non-spiral (NS) sub-samples, with
the  maximum  well correlated with the LOS direction. Another of the
studied directions may occur close to the maximum only if they are close
to the LOS and they do not correlate with other features visible  on
the  maps. For clusters with a clearly defined maximum of $s$ below
$3.0$ the conclusion generally does not  change.  For  the spiral (S)
sub-samples the maps are  usually at  the random  noise  level. In
these cases a weaker, but still existent correlation with the LOS is
observed and  no other evident correlations are noted. We conclude that
the strong systematic effect, generated by the process of galactic axis
de-projection from its optical image, is present in the catalogue data.
It can  mask  any  weak alignment existing in the analysed clusters.
With the use  of  a  simple model for the systematic effect  we  are
able  to  reproduce  the  main characteristic features of the maps for
non-spiral galaxies.  We  note, however, a few clusters showing
significant differences with respect  to this model.
\end{abstract}
 
\begin{keywords}
cosmology -- galaxy clusters -- galaxies: formation
\end{keywords}

\section{Introduction}
 
Some scenarios of cosmological structures evolution predict that the
orientation of galactic axes in clusters should prefer a certain direction
while in the other scenarios galaxies are expected to be randomly oriented
(cf. Shandarin 1974, Wesson 1982, Silk \& Efstathiou 1983, Dekel 1985).
Attempts to reveal any deviation from isotropy were performed over the
past years with different, and sometimes contradictory results. The work
done till 1985 is described by MacGillivray \& Dodd (1985), who point
out that most studies in that matter agree with the random galactic
distribution within the Local Supercluster (LSC) plane, however they
suggest that galactic planes can be oriented preferentially parallel
to the LSC plane at some distance from it. In an important paper
Kapranidis \& Sullivan (1983) analysed samples of bright spirals
belonging to the LSC and found no strong evidence for alignment of these
galaxies.
 
However, Jaaniste \& Saar (1977, 1978) claimed the existence of the mean
perpendicularity of galactic planes with respect to the LSC plane. Since
the earlier approaches were based mostly on analyses of highly inclined
and edge-on galaxies, Jaaniste \& Saar included all galaxies into
consideration, also the face-on ones. This approach was critically
discussed and modified by Flin \& God{\l}owski (1986). These authors and
later God{\l}owski (1993, 1994) analysed large galactic samples within
the LSC and came to the conclusion that there exists a preferential
orientation of galactic planes perpendicular to the LSC plane and that
there is evidence for aligning the galactic rotation axes along the
directions toward the Virgo cluster centre. The comparison of the
alignment properties of spiral and non-spiral galaxies reveals the
physically unexpected conclusion that spirals exhibit weaker alignment.
It occurs in spite of the fact that the alignment should result from the
correlated galactic angular momenta in the cluster and the spiral galaxies
carry substantial angular momenta allowing for more precise
determinations of its' spatial orientations to measure such correlations.
 
The recent study of Parnovsky et al. (1994) used both the analysis of
the galactic position angles and the directions of galactic axes
obtained from de-projections of their images in UGC and ESO catalogues.
Additionally, the catalogue of flat edge-on galaxies compiled by the
same authors was used. They detected an anisotropy with the galactic
axes distribution forming a three-axial ellipsoid, showing an excess of
about 20\% in the direction (4$^{\rm h}$-6$^{\rm h}$, 20$^\circ$ -
40$^\circ$), and a deficit of about 25\% in the direction (13$^{\rm
h}$-15$^{\rm h}$, 30$^\circ$-40$^\circ$). The results are in general
agreement with the earlier result of Fliche \& Souriau (1990) for
orientations of the extended H\,I galactic envelopes and the `cosmic
pole' at (5$^{\rm h}$30$^{\rm m}$, 7$^\circ$) derived from distant
quasars. Later discussion by Flin (1995) points out that the anisotropy
direction found by these authors is essentially the same as the result
of Flin \& God{\l}owski (1986). However, the detected galactic
orientation anisotropy is not global, but related to the LSC.
 
Several aspects of galactic orientations within separate clusters were
investigated in a few aspects. There seems to exist compelling evidence
for strong alignment of the CD galaxy with its cluster, and  the  effect
is stronger in more elongated clusters  (Struble  1987,  1990;  Mandzhos
1987; van Kampen \& Rhee 1990; Trevese et al. 1992, Han  et  al.  1995).
Muriel \& Lambas (1992) showed the existence of  systematic  effects  in
orientations of galaxies with respect to their  neighbours.  For  spiral
galaxies the effect exists with respect to the nearest neighbour,  while
for elliptical galaxies this is also true  for  all  neighbours  at  the
distances smaller than $3$ Mpc h$^{-1}$. The study  of  orientations  of
weaker galaxies in clusters (Trevese et al. 1992) does  not  reveal  any
significant alignment, confirming similar earlier reports. However,  the
reports of God{\l}owski (1993, 1994) based  on  the  analysis  involving
de-projection of galactic rotation  axes  (Flin  \&  God{\l}owski  1986)
reveal the possibility of the existence of some  systematic  differences
between different substructures  of  the  LSC  and  suggest  a  possible
`hedgehog-like' orientation pattern.
 
The work done so far, claiming the detection or non-detection of a galactic
alignment in galaxy clusters and other LSC substructures, with the
discrepant alignment characteristics, requires further study and
clarification. Since the large scale correlations between galactic
orientations can be decreased by mixing several differently oriented
sub-structures, the investigation of  separate  clusters  should  reveal
a much clearer picture of the preferred orientation, if any.
Therefore, the present work was originally intended to investigate the
galactic axes orientations within 18 selected clusters, substructures of
the Local Supercluster, based on data from the Tully's {\it Nearby
Galaxy Catalogue} (Section 2). For a majority of these clusters the number
of galaxies is not very large. For this reason we have decided to use
the statistical parameter of the Fourier test, $\Delta_{11}$,
describing the preferential orientation of galactic axes with respect to
the main axis of the cluster reference frame (Flin \& God{\l}owski 1986;
God{\l}owski 1993, 1994; see Appendix B). In Section 3, for every
investigated cluster, we map the value of $\Delta_{11}$ divided by its
formal error $\sigma(\Delta_{11})$, with the cluster pole co-ordinates
changing along the entire celestial sphere. The resulting maps are
analyzed for correlations of the $\Delta_{11} / \sigma(\Delta_{11})$
maxima with directions from the cluster centre to 1.) the derived
cluster poles, 2.) the Local Supercluster centre, 3.) the Virgo A centre
and 4.) the Earth, i.e. along the line of sight ($\equiv$ LOS). For any
cluster we divide the full sample according to morphological type into
spiral (S) and non-spiral (NS) sub-samples. The strong maxima of
$\Delta_{11} / \sigma(\Delta_{11})$, above $3.0$, with one exception
occur only for the NS sub-sample, where the maximum is well correlated
with the LOS direction. The maxima are correlated with other directions
when those directions are close to the LOS. These directions do not
correlate in a clear way with other features visible on the maps. For
structures with a clearly defined $\Delta_{11} / \sigma(\Delta_{11})$
maximum below $3.0$ the conclusion generally does not change. For the S
sub-samples the maps are often at the noise level  with  the  maxima  of
$\Delta_{11} / \sigma(\Delta_{11}) \approx 1.0$. In these cases no clear
correlation of the considered directions and features  on  maps  may  be
seen. We conclude that the strong systematic effect,  generated  by  the
process of galactic  axis  de-projection  from  its  optical  image,  is
present in the catalogue data. It can mask any existing  weak  alignment
in the analysed clusters. The divergences from the expected form of  the
map observed in some cases may  indicate  the  existence  of  non-random
galactic  distributions.  The  effects  however  are  too  weak  to   be
considered quantitatively with the present method. In  section  4,  with
the use of a simple model for this systematic effect,  one  is  able  to
reproduce the main characteristic features of the  maps  for  non-spiral
galaxies. A short summary and discussion is given in  the  last  section
(Section 5). The present analysis provides a firm argument, that one
should be very careful when using
of methods involving reproductions  of  galactic  axes  orientations
from the shapes of  their  images  in  statistical  investigations.  The
available data for non-spiral galaxies are subject to  large  systematic
errors. The use of such data  in  the  alignment  analysis  for  samples
distributed over large parts of the sky could lead to the  detection  of
spurious alignments toward regions containing  the  highest  numbers  of
galaxies. The analysis of galactic  position  angles  is  free  of  such
systematic effects.
 
\section{OBSERVATIONAL DATA}
 
We considered a galactic sample selected from Tully's (1988) {\it Nearby
Galaxy Catalogue}, including LSC  galaxies  with  radial  velocities  --
corrected for the solar motion  --  smaller  then  $2800$  km/sec.  This
catalogue provides estimates for galactic  distances,  group  membership
and the derived galactic axis  inclination  to  the  LOS.  The  galactic
position angles necessary in our analysis are not available  in  Tully's
catalogue and were therefore taken from Nilson (1973, 1974) and  Lauberts
(1982, 1989). For about $10$  percent  of  the  considered  objects  the
position angels are not available in any  source.  However,  in  general
they are the nearly face-on galaxies and  their  position  angle  values
have only negligible effects on the derived galactic axes  orientations.
For the numerical computations below we have taken these position angles
at random from the uniform distribution. Tully claims that the data from
his catalogue is free of the Holmberg effect, thus no further correction
for that effect is introduced into the analysis.
 
It is well known that the galactic brightness is a very  poor  distance
indicator for neighbour galaxies. For this reason Kapranidis \& Sullivan
(1983) in their study of galactic  alignment  obtained large  differences
between samples selected according to radial velocities  and  brightness
(see also Flin \& God{\l}owski 1986). Therefore knowledge of  the  group
membership derived from radial  velocities  is  crucial  for  a  serious
analysis. Under present  considerations  we have selected  such  defined
clusters of galaxies from Tully's catalogue. We have taken  into  account
all  clusters  which  consist  of  at  least  40  galaxies.  We analysed
separately  the  samples  including  all  cluster   galaxies   and   two
sub-samples selected according to the galactic  morphological  type.  We
extracted spiral galaxies as one sub-sample (S) and  non-spiral,  mostly
elliptical ones as the second sub-sample (NS). The  considered  clusters
are labeled with Tully's numbers, while the names of the  widely  known
ones are given in parentheses: 11 (Virgo), 12 (Ursa Major),  13,  14
(Coma), 15, 17, 21 (Leo), 22, 23, 31  (Antila-Hydra),  41,  42,  44,  51
(Fornax-Eridianus),  52,  53  (Dorado),   61   (Telescopium),   64.   An
orientation sketch of the Local Supercluster of galaxies with  indicated
positions of the discussed galactic clusters --  sub-structures  of  the
LSC -- is presented in Fig.~1. Detailed lists  of the  galaxies  in  the
investigated groups can be obtained upon request from one of  the  authors
(WG).
 
In studies following the Jaaniste \& Saar (1977) approach a crucial
point is to obtain a correct value for the galaxy inclination angle $i$.
A formula enabling derivation of that angle from the observed axial
ratio $q$, valid for oblate spheroids, is provided by Holmberg (1946) as
 
$$\cos^2{i}=(q^2-q_0^2)/(1-q_0^2) \qquad . \eqno(2.1)$$
 
\noindent
Tully (1988) in the NGC catalogue used this formula with $q_0 = 0.2$ for
obtaining inclination angles.  Only  in  a  statistically  insignificant
number of cases, he obtained inclinations from other  information,  such
as the form of  rings  or  spiral  structure.  Then,  he  increased  the
obtained values of $i$ by $3^\circ$, in accordance  with  the  empirical
recipe given by Aaronsen et al. (1980). Let us note that Kapranidis  \&
Sullivan (1983)  used  such  a  formula  (without  the  mentioned recipe
`$+3^\circ$') only for spirals. Heidmann et al. (1971) noticed that  the
value  of  $q_0$  depends  on  the  galactic  morphological   type   and
consequently should vary for different types of  galaxies:  for  spirals
$q_0$ should be less than $0.2$,  while  for  elliptical  galaxies  they
should be significantly higher. For  this  reason  Flin  \&  God{\l}owski
(1986) and later God{\l}owski (1993, 1994) considered the  `true'  axial
ratio $q_0$ to depend on the galaxy morphological type. However,  as  we
argue later, it may be questioned if such simplified approach  may  work
reasonably well.
 
\begin{figure*}
\vskip 12cm
\includegraphics{siat.ps}
\caption{An orientation sketch of the  Local  Supercluster  of  galaxies
with the positions of galaxy clusters used in  our  analysis  indicated.
The supergalactic co-ordinates ($X$, $Y$, $Z$) are  given.  The  cluster
numbers are given near symbols indicating galaxy clusters  and  straight
lines join these symbols with their projections on the  LSC  plane.  Our
position is marked with `E'.}
\end{figure*}

\section{Testing the galactic orientations in clusters}
 
\subsection{The data analysis method}
 
Let us shortly summarize the data analysis method of Flin \&
God{\l}owski (1986; see also God{\l}owski 1993, 1994) used for the
investigation of possible galaxy alignments in LSC clusters. In the
analysis one uses the supergalactic co-ordinate system ($L$, $B$, $P$)
with the basic great circle (`meridian') chosen to pass through the LSC
centre in the Virgo cluster. For any galaxy we consider two parameters:
the galactic position angle $p$ and the inclination angle $i$. With the
use of these angels two orientation angles are determined: $\delta$ - an
angle between the normal to the galaxy and the LSC plane, and $\eta$ -
an angle between the projection of this normal at the LSC plane and the
direction toward the LSC centre. In the present analysis the method was
applied for angles defined with respect to the individual galaxy cluster
main plane instead of the LSC plane. The distributions of the two angles
$\delta$ and $\eta$ can be analyzed using statistical tests from Flin \&
God{\l}owski (1986) shortly summarized in Appendix B. Within this method
based on de-projection of galactic images one obtains two possible
vectors normal to the galactic plane and both solutions are considered
in the analysis (Jaaniste \& Saar 1977). The method was previously used
by one of us to study the galactic alignment inside the whole LSC.
However, when we consider individual clusters the number of galaxies
involved may be small in some cases and not all statistical tests
described by Flin \& God{\l}owski (1986) will work well (e.g. $\chi^2$
test requires the expected number of data per bin to equal at least 7;
see, however, Snedecor \& Cochran 1967 and Domanski 1979). We base our
present work on a test involving the Fourier coefficient $\Delta_{11}$
(cf. Appendix B). The value of $\Delta_{11}$ characterizes the mean
alignment of galactic planes with respect to the chosen axis of the
reference frame. The positive value of $\Delta_{11}$ obtained during the
analysis of the angle $\delta$ appears when the galactic rotation axes
tend to be perpendicular to the cluster plane, and for the negative
value these axes are preferentially parallel to that plane\footnote{If
one calculates $\Delta_{11}$ for the angle $\eta$, the value of
$\Delta_{11} < 0$ occurs when the projection of the galactic rotation
axis at the cluster's plane is preferentially oriented perpendicular to
the zero point direction of $\eta$, and for $\Delta_{11} < 0$ the
projection tends to be directed along this direction.}.
 
As described in Appendix A, the galactic cluster main plane and the
respective axis perpendicular to the cluster main plane can be derived
by assuming that the galactic cluster is a three-axial spheroid with
Gaussian density distribution along each axis. We establish directions
of these axes in the considered cluster by fitting the three dimensional
the galactic distribution to galactic positions taken from Tully's
catalogue. As the actual clusters are often quite irregular and subject
to inevitable galactic distance errors the derived axes are rather
formal ones and only sometimes contain physical information about the
cluster gravitational potential structure. In the analysis below, we
allow for all possible orientations of the cluster's main axis along the
celestial sphere. Of course, this axis is not related to the previously
mentioned `physical' axes. For each particular choice of the cluster
pole the value of the Fourier parameter $\Delta_{11}$ for the angle
$\delta$ is derived in order to seek the maximum value of $| \Delta_{11}
|$. Because we change the `pole' position along the whole sphere,  there
is no need to repeat the analysis for the  angle  $\eta$.  Maps  of  the
obtained $\Delta_{11}$ are presented in supergalactic  co-ordinates  $L$
and $B$ in Fig.~2. More precisely, we plot the value of  $  \Delta_{11}$
divided by its standard deviation  $\sigma(\Delta_{11})$  ($\Delta_{11}/
\sigma(\Delta_{11})  \equiv  s$).  It  is  sufficient  to  present   one
hemisphere only,  the  second  one  is  obtained  reflecting  oppositely
directed poles. Next to each map  the  considered  cluster  number,  the
number of considered galaxies and the maximum value of $| \Delta_{11}  /
\sigma(\Delta_{11}) |$ in the map are given. For S and NS galaxies, the
respective symbol is positioned after the indicated number of galaxies.
For presentation in Fig.~2 we selected four clusters (11, 15, 22, 23)
with different characteristic features, the full set of maps for 18
clusters is provided in Appendix C. For each cluster, maps are provided
for all galaxies and for both S and NS sub-samples. The maps are
presented in the order of decreasing maximum value of $| s |$. On each
map we denote the crucial directions, as seen from the centre or toward
the centre \footnote{We chose one of the symmetric directions seen on
the presented hemisphere.} of the considered cluster: 1.) the three
derived cluster poles (cf. Appendix A), 2.) the direction to the Local
Supercluster centre, 3.) the direction of the Virgo cluster  centre  and
4.) the line of sight  to  the  Earth  ($\equiv  LOS$).  Data  from  the
respective fits for all 18 clusters is summarized in Table~1,  including
the cluster number according to Tully, numbers of ALL, S and NS galaxies
in the cluster, position of the Earth as seen from the cluster centre
($L_E$, $B_E$) and its distance $R$ in Mpc (assuming $H_0 = 75$
km/s/Mpc), and the position of the parameter $s$ maximum on the map for
ALL galaxies, ($L_{max}$, $B_{max}$), followed by the corresponding
value of that parameter.
 
\renewcommand{\thefigure}{\arabic{section}a}
\setcounter{section}{2}
 
\begin{figure*}
\vskip 22cm
\includegraphics{surs1.ps}
\caption{ Cluster 11, Cluster 15 }
 \end{figure*}
 
\renewcommand{\thefigure}{\arabic{section}b}
 
\begin{figure*}
\vskip 22cm
\includegraphics{surs2.ps}
\caption{ Cluster 22, Cluster 23}
\end{figure*}
 
\renewcommand{\thefigure}{\arabic{section}}
\setcounter{section}{2}
 
\begin{figure*}
\vskip 0cm
\caption{Maps of $s \equiv \Delta_{11} / \sigma(\Delta_{11})$ versus the
chosen cluster pole supergalactic co-ordinates ($L$, $B$). On the maps,
the results of the real data analysis are shown on the left, while the
maps obtained from our systematic effect modelling (Section 4) are given
on the right. For each cluster the maps are presented for ALL cluster
galaxies and for S and NS sub-samples. On the map we indicate the
important directions, as seen from the centre of the considered cluster:
1.) three cluster poles - see text (full star, square and triangle), 2.)
the direction to the Local Supercluster centre (open circle), 3.) the
direction of the Virgo A cluster centre (open square) and 4.) the line
of sight from the Earth (asterisk). Near each panel we give the cluster
number appended for the sub-samples with the respective symbol S or NS, the
number of galaxies in the (sub-) sample $N$ and the maximum value of $s$
on the map. Four examples are presented in Fig.2: a.) Cluster 11,
and Cluster 15, b.) Cluster 22 and Cluster 23 .} \end{figure*}
 
\setcounter{section}{3}
\subsection{Analysis Results}
 
In the majority of the considered clusters we observe two `hills' of
positive values in opposite directions on the celestial sphere
separated by a circular `valley' of negative values (see Fig.~4 in
Appendix C). The positive hill is expected to appear on the map if the
direction of the galactic cluster axis is close to the direction of
galactic axes alignment. If we take the considered axis perpendicular to
that direction, the alignment is mostly perpendicular to the axis and we
observe negative values of $\Delta_{11}$. For the alignment parallel to
a single axis in the cluster the depth of the negative valley on the
map should be smaller than the height of the hills. In the case when the
galactic axes align uniformly along the plane, one expects to find
negative holes in the directions perpendicular to that plane, separated
by a somewhat less pronounced positive `bank' for the cluster axis
directions chosen in the considered plane. Thus the structure most often
observed (see cluster 11 at Fig.~2a, Fig.~4) could be produced if the
galactic axes tend to align along the line joining the considered hills
and then the observed alignment effect could often be interpreted as
statistically  significant.  Below  we  will  argue  against   such   an
interpretation and we suggest that a systematic effect along the line of
sight may be to a large extent responsible for the structures visible on
the maps. One should note, however, that the picture with positive hills
surrounded by negative valleys is not a strict rule for all clusters. In
some statistically less-significant cases, like cluster 23 in Fig.~2b,
one has separate negative minima surrounded by a circle of positive
values. Sometimes the relative values of positive and negative
maxima do not fit the above described picture well (for example cluster
52 in Appendix C with nearly equal amplitudes). For cluster 15, on
the map for ALL galaxies (Fig.~2a) the observed maximum alignment
direction does not correlate with the LOS and in cluster 22 we do
not see any significant peaks.
 
A significant difference is claimed to exist in alignment
characteristics of the spiral galaxies and non-spirals ones (e.g.
Flin \& God{\l}owski 1986). Therefore, in the next step of the analysis
we consider differences between such sub-samples in each cluster
(Fig.~2, Fig.~4). For NS galaxies one generally observes a positive peak
strongly correlated with the LOS direction, surrounded by the valley of
negative $s$~. An  exception  is  cluster  No.~52,  where  a  pronounced
negative `peak' is encountered. For spiral  galaxies,  a  different
picture is common. In general  one  does not observe the above 3  sigma
alignment effects (i.e. with $s > 3.0$). Usually the obtained picture is
consistent with the random distribution appended with a very weak
systematic correction along the LOS. An exception is cluster No.~15,
where we observe a pronounced positive peak far from the LOS, which
could be an indicator of real alignment.
 
In Fig.~2 and Fig.~4, the evident strong correlation of the positive
maximum with the LOS direction exists for NS galaxies. It is clear
that the same correlation on the maps for ALL cluster galaxies is to a
large extent generated by the NS component. In addition, because of the
special position of the Earth with respect to the investigated clusters
the LOS and the Virgo cluster direction are often close to each other
(cf. Fig.~1). In the considered clusters, a coincidence between the Virgo
direction and alignment maximum occurs only if the Virgo direction is
close to the LOS, and it never happens if it is far from the LOS. In a
few cases where the Virgo direction is close to this maximum, the LOS is
generally somewhat closer. This fact suggests that the systematic error
in determining the galactic axis inclination angle is present in Tully's
catalogue which was used for this analysis. The effect -- turning galactic
`faces' toward the observer to produce the observed `alignment' -- is
much more pronounced for NS galaxies, but can also be detected on the
maps for the S ones. The last statement can be checked by inspecting of
all maps presented in Appendix C (Fig.~4) for the S sub-samples. In the
following section we propose a simple model which allows to reproduce
the characteristic systematic effects seen on the maps.
 
\begin{table*}
\caption{A list of analysed clusters: the respective numbers ($N$,
$N_S$, $N_{NS}$), cluster positions ($L_c$, $B_c$, $R$), the LOS
position at the presented map ($L_E$,$B_E$), the maximum $s$ position at
the map ($L_{max}$, $B_{max}$), and the value of this maximum }
 
\begin{tabular}{crrrrrrrrrrr}
Cluster&
\multicolumn{3}{c}{Object No}&
\multicolumn{3}{c}{Position}&
\multicolumn{2}{c}{LOS}&
\multicolumn{3}{c}{Maximum}\\
No&
\multicolumn{1}{c}{$N$}&
\multicolumn{1}{c}{$N_S$}&
\multicolumn{1}{c}{$N_{NS}$}&
\multicolumn{1}{c}{$L_c$}&
\multicolumn{1}{c}{$B_c$}&
\multicolumn{1}{c}{$R$}&
\multicolumn{1}{c}{$L_E$}&
\multicolumn{1}{c}{$B_E$}&
\multicolumn{1}{c}{$L_{max}$}&
\multicolumn{1}{c}{$B_{max}$}&
\multicolumn{1}{c}{$s$}\\
$11$&$313$&$227$&$ 86$&$ 11.7$&$ -2.6$&$22.4$&$191.7$&$ 2.6$&$186.0$&$
2.0$&$6.1$\\
$12$&$166$&$123$&$ 43$&$311.1$&$ 3.0$&$22.1$&$311.1$&$ 3.0$&$310.0$&$
5.0$&$3.1$\\
$13$&$ 64$&$ 38$&$ 26$&$320.9$&$ -7.1$&$28.3$&$140.9$&$ 7.1$&$350.0$&$
5.0$&$2.2$\\
$14$&$213$&$101$&$112$&$357.5$&$ 2.8$&$ 5.2$&$357.5$&$ 2.8$&$150.0$&$
3.0$&$2.8$\\
$15$&$ 65$&$ 47$&$ 18$&$327.9$&$-23.3$&$ 8.8$&$147.9$&$ 23.3$&$
59.0$&$46.0$&$2.6$\\
$17$&$ 40$&$ 23$&$ 17$&$225.5$&$ -7.3$&$10.4$&$ 45.5$&$ 7.3$&$ 50.0$&$
5.0$&$1.8$\\
$21$&$124$&$ 90$&$ 34$&$326.6$&$-23.6$&$24.5$&$146.6$&$
23.6$&$147.0$&$15.0$&$2.6$\\
$22$&$ 63$&$ 40$&$ 23$&$ 25.7$&$-19.2$&$26.1$&$205.7$&$
19.2$&$170.0$&$58.0$&$0.9$\\
$23$&$ 50$&$ 37$&$ 13$&$ 59.1$&$ -9.2$&$37.4$&$239.1$&$ 9.2$&$ 60.0$&$
5.0$&$1.5$\\
$31$&$105$&$ 79$&$ 26$&$ 30.8$&$-46.6$&$32.0$&$210.8$&$
46.6$&$210.0$&$45.0$&$3.3$\\
$41$&$ 96$&$ 63$&$ 33$&$ 22.6$&$ 24.0$&$29.7$&$ 22.6$&$ 24.0$&$
30.0$&$27.0$&$3.6$\\
$42$&$115$&$ 82$&$ 33$&$313.2$&$ 18.5$&$35.0$&$313.2$&$
18.5$&$312.0$&$12.0$&$2.1$\\
$44$&$ 40$&$ 26$&$ 14$&$307.7$&$ 35.0$&$17.9$&$307.7$&$
35.0$&$313.0$&$15.0$&$2.1$\\
$51$&$114$&$ 80$&$ 34$&$174.8$&$-40.3$&$19.8$&$354.8$&$ 40.3$&$
1.0$&$45.0$&$4.1$\\
$52$&$ 86$&$ 64$&$ 22$&$198.2$&$-17.0$&$22.2$&$ 18.2$&$ 17.0$&$
30.0$&$12.0$&$3.7$\\
$53$&$130$&$ 91$&$ 39$&$132.6$&$-44.4$&$13.3$&$312.6$&$ 44.4$&$
31.0$&$33.0$&$1.2$\\
$61$&$129$&$ 99$&$ 30$&$138.4$&$ 12.3$&$25.7$&$138.4$&$
12.3$&$170.0$&$30.0$&$2.0$\\
$64$&$ 51$&$ 33$&$ 18$&$199.9$&$ 29.8$&$26.9$&$199.9$&$
29.8$&$203.0$&$30.0$&$3.5$
\end{tabular}
\end{table*}

\setcounter{section}{3}
 
\section{Modelling the considered systematic effect}
 
In order to quantitatively evaluate the considered systematic effect we
have performed a simple modelling of the anisotropic galactic
distribution in order to reproduce Tully's catalogue data. Our null
hypothesis is that in fact we have an isotropic distribution of galaxy
angular momenta and any preferred orientation arises due to favouring
the LOS direction in elaboration of the observational data. This fact
suggests that the effect should be expressed as a function of  the
galaxy inclination angle cosine $\mu$. Thus, we proceeded in the
following way. For any considered cluster the galactic spatial positions
were not modified. We considered the same distribution of galactic poles
for all clusters, consisting of an isotropic part and a simple
anisotropic correction along the LOS, proportional to $\mu$ (i.e. we
consider only the first two terms of the distribution expansion in
powers of $\mu$, an example with included terms $\propto \mu^2$ is
analyzed in Appendix D). The assumed normalized probability distribution
has the form
 
$$F(\mu) = {1 \over 1+A} + {2 A \over 1+A}\, \mu \qquad . \eqno(4.1)$$
 
\noindent
$\mu$ is the cosine of the angle between the LOS and the galactic pole
and the distribution exhibits rotational symmetry along the LOS. For
each galaxy in the analysis one uses both solutions with the positive
and the negative value of $\mu$, but, due to symmetry, the range $0 <
\mu < 1$ is considered in Eq.~4.1~. The anisotropic part has a simple
form with the (positive or negative) amplitude parameter $A$.
 
If the effect of anisotropy is systematic it should influence all
galaxies in the catalogue, not only those contained in the discussed
clusters. Therefore, for determination of the amplitude $A$ we
considered all LSC galaxies from the Tully's catalogue. ALL, S and NS
galaxies were considered separately and the derived amplitude $A$ is
appended with the respective index $ALL$, $S$ or $NS$. Let us note that
with thus obtained $A$ it was possible to derive the systematic effect
for the whole LSC to have the similar $\Delta_{11}$ as the value derived
from the original data. For ALL galaxies the best fit gives
 
$$ A_{ALL} = 0.7 \qquad . \eqno(4.2)$$
 
\noindent
An integral amplitude of the anisotropic part is $$\int^1_0 2A_{ALL} \mu
/ (1+A_{ALL}) d\mu = 0.41 \quad .$$ A fit quality of the data from the
Tully catalogue for ALL LSC galaxies can be evaluated by an inspection of
Fig.~3 for both galactic orientation angles $\delta$ and $\eta$. In the
figure, we presented the catalogue data superimposed over the `theoretical'
distribution derived as a mean from 1000 simulated catalogues using the
distribution (4.1) with the amplitude $A$ value given in (4.2).
 
\renewcommand{\thefigure}{\arabic{section}}
\setcounter{section}{3}
\begin{figure}
\vskip 11.5cm
\includegraphics{fi.ps}
\caption{ A fit of the real data to the theoretical model for ALL
galaxies in the LSC. The fitted angle is: a.) $\delta$ , b.) $\eta$.
For each galaxy both possible orientations of the axis (`solutions') are
included.}
\end{figure}
The discussion in the previous section notes a significant difference
between distributions for spiral and non-spiral galaxies. For the spiral
galaxies our fitting procedure yields a much smaller value of the
amplitude $A$
 
$$A_S = 0.15 \qquad , \eqno(4.3)$$
 
\noindent
while for the non-spirals
$$A_{NS} = 20.0 \qquad . \eqno(4.4)$$
 
\noindent
One should remember that for the spiral galaxies some low brightness
face-on spirals may be missing and the obtained $A_S$ may be
underestimated.
 
With the use of the derived average values for $A$ (4.2-4), in Fig.~2
and in Appendix C we plotted the maps obtained for the considered
clusters with the model distribution (4.1), to be compared with the real
data. In each case we derived only one model map, and this map --
including the statistical fluctuations -- is presented. One may note
that our simple model reproduces the original data for clusters with the
large maxima of $\Delta_{11} / \sigma (\Delta_{11})$ relatively well
and thus provides a strong argument for the presence of the suggested
systematic effect. For spiral galaxies the maps are often at the one
sigma level, but quite often the simulated map is similar to the real
one with the LOS close to the maximum on the map. However, in some cases
noticeable deviations between the simulated and the real maps exist. In
some clusters the expected significant maximum along the LOS does not
exist (cf. Cluster 22 or 15 at Fig.~2). Also the depth of a negative
valley surrounding the maximum changes in some cases significant in more
than expected for the uniform and isotropic galactic distribution
and observed maximum is much higher then expected from the model
(cf NS Cluster 11). In our opinion, the obtained large value of $A_{NS}$
proves that Tully's procedure for derivation of the galactic axis
inclination provides values not entirely related to the real galactic
orientations. Probably his use of a mean galactic ellipsoid form for
such galaxies is in appropriate.
 
Finally, we would like to stress that the presented model is not
expected to reproduce all details seen on the maps. Beside the
statistical fluctuations, the observed deviations between real and model
maps can reflect not only the actual galactic alignments, but also the
fact that the systematic error in the catalogue may have a somewhat
different form than the simple test distribution introduced in the
equation (4.1) and/or it can arise from fine galactic morphological type
variations in individual clusters, leading to a slightly different
systematic effects. An example of a more thorough analysis of the
problem is presented in Appendix D for the Virgo cluster.
 
\setcounter{section}{4}
 
\section{Final remarks}
 
We studied the galactic axes alignments in 18 clusters from the LSC with
the method based on the analysis of de-projected galactic axes
orientations. The considered statistical parameter $\Delta_{11}$
describes the galactic axes alignment along the given direction. Mapping
of this parameter in all directions on the sky shows that the most
evident alignment occurs along the line of sight, at least in the
majority of statistically significant cases. We interpret this finding
as the result of the systematic error of the galactic axis determination
in the Tully (1988) catalogue.
 
The anisotropy modelling of the previous section shows  the  possibility
of obtaining a reasonable fit to the data with the  simple  model  (4.1)
and the same value of the anisotropy parameter $A$ for all clusters. One
may note that the introduced anisotropy is proportional to the  galactic
axis projection on the  LOS,  i.e.  the  derived  galactic  distribution
underestimates the number of edge-on galaxies. In Tully's catalogue this
effect is very strong for non-spiral galaxies ($A_{NS} =  20$),  but  it
can also be noticed in the spiral galaxies'  sub-sample,  where  $A_S  =
0.15$.  The  analysis  of  the  maps  for  spiral  galaxies,  where  the
systematic effects  are  much  weaker,  do  not  reveal  any  meaningful
correlation of the considered directions -- other than the LOS  --  with
the maps' characteristic features. Possible problems with  de-projection
of elliptical  galaxies  was  noted  by  some  authors  previously  (cf.
Kapranidis \& Sullivan 1983). In the present paper we provide convincing
proof for such an effect to exist in Tully's catalogue. In  our  opinion
the discussed LOS effect arises during computation  of  the  inclination
angle, where Tully takes the `true' galaxy axial ratio  equal  to  $0.2$
for all galaxies, independently of their morphological  type\footnote{We
also performed (God{\l}owski, in preparation) the analysis  of  galactic
orientations in the UGC and the ESO catalogues  (Nilson  1973,  Lauberts
1982). In order to derive galactic spatial orientations, the Heidmann  et
al. (1971) recipe for the galaxy shape determination  and  the  standard
Holmberg formula for the inclination angle were used.  The  model  (4.1)
was fitted to thus obtained  data.  The  obtained  values  of  $A$  were
significantly smaller than  the  ones  given  in  (4.2-4).  This  result
confirm our conclusion that the de-projection procedure  is  responsible
for the observed systematic effect.}.
 
Let us mention that opposed to our finding Bahcall et al. (1990) claim
that no excess of face-on ellipticals and the excess of edge-on spirals
exist for the UGC galaxies. The last effect arises probably due to the
number of missing face-on spirals.  Our  positive  fit  for  $A_S  >  0$
suggests that either there are very few missing face-on spirals  in  the
Tully catalogue, or the systematic de-projection errors are  significant
enough for spiral galaxies to compensate for the lack  of  some  face-on
galaxies (we do not consider the possibility of a real  alignment  along
the elongated structure of the LSC here).
 
Let us mention that some previous positive detections of galactic axes
alignment can be dominated by similar systematic effects introduced by
the methods involving de-projection of galactic images. In particular,
in our analysis, the LOS effects are much larger than any actual signal,
yielding a spurious galactic alignment with the direction to the
cluster. When analyzing larger samples covering a substantial part of
the sky -- with objects distributed in a non-uniform way -- the effect
of this type should introduce galactic axes alignment along the
direction pointing toward the region containing the largest
concentration of galaxies. In a similar manner, a deficit of alignment
would be produced in the directions of less populated parts of the sky.
In our opinion there is no safe approach, excluding the possibility of
the LOS systematic effects, when attempting to derive galactic spatial
orientations for large galactic samples. The alignment studies involving
the galactic position angles should not be influenced by any systematic
effect of the kind discussed in the present paper if the gravitational
lensing does not play a role.
 
Of course, our analysis does not rule out the possibility of some real
weak galactic alignments within galaxy clusters. The exceptions from the
general picture involving the systematic effect mentioned in the text
provide some weak evidence for the real alignments existing in a
few clusters. For example the pronounced positive peak for spiral
galaxies occurring far from the LOS in Cluster 15 may be suspected to
represent such a real alignment, see also the discussion of positive and
negative maxima in section 3. However, one should be aware of the fact
that the statistical significance of any such exception may be much
lower than the value $\Delta_{11} / \sigma(\Delta_{11})$ provided at the
map due to the large number of tested directions, i.e. a large number of
statistical trials.
 
\section*{Acknowledgement}
 
WG thanks Dr. Frank Baier for our discussions. We are grateful
to the anonymous referee for suggestions which helped to considerably
improve the paper.
 
\section*{Appendix A. Fitting of the galaxy cluster axis}
 
We use the supergalactic co-ordinate system ($L$, $B$, $P$), where $P$
is the supergalactic position angle defined by Flin \& God{\l}owski
(1986). The co-ordinates of the supergalactic pole in the equatorial
system $\alpha_{1950} = 285.5$ and $\delta_{1950} = +16$ are taken from
Tammann \& Sandage (1976). In our system the basic great circle
(`meridian') of the supergalactic system is chosen in such a way that it
passes through the Virgo cluster centre with co-ordinates $\alpha_{1950}
= 186.25$, $\delta_{1950} = +13.1$ (Tammann \& Sandage 1976). In Tully's
(1988)  catalogue,  galactic  distances  and  their  membership  in  the
corresponding groups are given. With the use  of  this  data  we  derive
positions for cluster's galaxies in an orthogonal  reference  frame.  To
the obtained  distribution  we  fit  the  triaxial  ellipsoid  with  the
Gaussian   density   distribution   along   each   axis.   Through   the
orthogonalization procedure of the covariance  matrix  of  a  considered
distribution,  we  derive  the  directions  of  the  main  axes  of  the
considered cluster.
 
\section*{Appendix B. The applied statistical procedures}
 
To check the distribution of galactic orientation angles ($\delta,
\eta$) we applied statistical tests originally introduced to this
problem by Hawley \& Peebles (1975), later improved by Kindl (1987) and
described in detail by God{\l}owski (1993, 1994). Below, a short summary
is presented of the $\chi^2$-test, the Fourier test and the
autocorrelation test.
 
Let $N$ denote the total number of solutions for the galactic axes
(two solutions for any galaxy) in a considered cluster, $N_k$ -- the
number of galaxies with orientations within the $k$-th angular bin,
$N_0$ -- the mean number of galaxies per bin and, finally, $N_{0,
k}$ -- the expected number of galaxies in the $k$-th bin. In the present
derivations we adopted, in most cases, a division for $n = 36$ bins of
equal width. As a check, in a few cases we repeated the derivations for
different values of $n$ but no significant differences appeared.
 
The presented in section 4 fits of the anisotropy parameter $A$ for all
LSC galaxies were  obtained  with  the  use  of  the  statistical  tests
described below. The $\chi^2$-test  of  the  distribution  involves  the
value
 
$$\chi^2 = \sum_{k = 1}^n {(N_k -N_{0,k})^2 \over N_{0,k}} \qquad .
\eqno(B1)$$
 
\noindent
It is a subject of the minimalization in model fitting, with the use of the
distribution provided by the model, $N_{0,k}$. In the present paper this
test was used to fitting the parameters $A$ from all LSC galaxies and in
Appendix D for the analogous fit for the Virgo cluster. The main
statistical test used in the present paper is the Fourier test involving
the first Fourier mode only. The actual distribution $N_k$ is
approximated by
 
$$N_k = N_{0,k} (1+\Delta_{11} \cos{2 \theta_k} +\Delta_{21} \sin{2
\theta_k} ) \qquad . \eqno(B2)$$
 
\noindent
The coefficients $\Delta_{i1}$ ($i = 1, 2$) are given as:
 
$$\Delta_{11} = {\sum_{k = 1}^n (N_k -N_{0,k})\cos{2 \theta_k} \over
\sum_{k = 1}^n N_{0,k} \cos^2{2 \theta_k} } \qquad , \eqno(B3)$$
 
$$\Delta_{21} = { \sum_{k = 1}^n (N_k-N_{0,k})\sin{2 \theta_k} \over
\sum_{k = 1}^n N_{0,k} \sin^2{2 \theta_k} } \qquad , \eqno(B4)$$
 
\noindent
with the standard deviation
 
$$ \sigma(\Delta_{11}) = \left( {\sum_{k = 1}^n N_{0,k} \cos^2{2 \theta_k}
} \right)^{-1/2} = \left( {2 \over n N_0} \right)^{1/2} \qquad ,
\eqno(B5a)$$
 
$$ \sigma(\Delta_{21}) = \left( {\sum_{k = 1}^n N_{0,k} \sin^2{2 \theta_k}
} \right)^{-1/2} = \left( {2 \over n N_0} \right)^{1/2} \qquad .
\eqno(B5b)$$
 
\noindent
The probability of the amplitude
 
$$\Delta_1 = \left( \Delta_{11}^2 + \Delta_{21}^2 \right)^{1/2}
\eqno(B6)$$
 
\noindent
being greater than a certain chosen value is given by the formula:
 
$$P(>\Delta_1 ) = \exp{\left( -{n \over 4} N_0 \Delta_1^2 \right)}
\qquad , \eqno(B7)$$
 
\noindent
while the standard deviation of this amplitude is
 
$$ \sigma(\Delta_1) = \left( {2 \over n N_0} \right)^{1/2} \qquad .
\eqno(B8)$$
 
\noindent
From the value of $\Delta_{11}$ one can deduce the direction of the
departure from isotropy. If $\Delta_{11} < 0$, then, for $\theta \equiv
\delta + \pi/2$, an excess of galaxies with rotation axes parallel to
the cluster plane is present. For $\Delta_{11} > 0$ the rotation axes
tend to be perpendicular to the cluster plane.

\section*{Appendix C. Maps of the parameter $s$ for 18 clusters}
 
\noindent
The maps of $s \equiv \Delta_{11} / \sigma(\Delta_{11})$ in
supergalactic co-ordinates ($L$, $B$) are presented for all considered
clusters (Fig.~4). For each cluster, the maps for
real data are presented above the respective maps obtained from our
systematic effect modelling. The maps are
given for ALL cluster galaxies and for S and NS sub-samples. On each
map we indicated the important directions, as seen from the centre of the
considered cluster: 1.) three cluster poles (full star, square and
triangle), 2.) the direction to the LSC centre (open circle), 3.) the
direction of the Virgo A cluster centre (open square) and 4.) the LOS
(asterisk). Near each panel we give the cluster number appended for
sub-samples with the respective symbol S or NS, the number of galaxies
in the (sub-) sample $N$ and the maximum value of $s$ on the map.
 
\renewcommand{\thefigure}{\arabic{section}a}
\setcounter{section}{4}
 
\begin{figure*}
\vskip 22cm
\includegraphics{su1.ps}
\caption{Clusters 11, 12, 13, 14 }
\end{figure*}
 
\renewcommand{\thefigure}{\arabic{section}b}
 
\begin{figure*}
\vskip 22cm
\includegraphics{su2.ps}
\caption{Clusters 15, 17, 21, 22 }
\end{figure*}
 
\renewcommand{\thefigure}{\arabic{section}c}
 
\begin{figure*}
\vskip 22cm
\includegraphics{su3.ps}
\caption{Clusters 23, 31, 41, 42 }
\end{figure*}
 
\renewcommand{\thefigure}{\arabic{section}d}
 
\begin{figure*}
\vskip 22cm
\includegraphics{su4.ps}
\caption{Clusters 44, 51, 52, 53}
\end{figure*}
 
\renewcommand{\thefigure}{\arabic{section}e}
 
\begin{figure*}
\vskip 12cm
\includegraphics{su5.ps}
\caption{Clusters 61, 64 }
\end{figure*}

\section*{Appendix D. Differences between the real map and the model}
 
As an interesting exercise one could apply the discussed  model  to  any
particular cluster in order to remove the influence  of  the  systematic
effect. Of course galaxies of different morphological types are included
in varying proportions in individual clusters and  therefore  the  model
anisotropy amplitude parameter $A$ may vary among them,  deviating  from
the mean value derived by us. Also, the simplified  model  claiming that
the systematic effect is simply due to  the  term  proportional  to  the
galaxy inclination
cosine may be a substantial oversimplification in some clusters. Such
effects are visible by eye among the maps of Fig.~4. However, as even a
very imperfect removing of the LOS effect will not erase the off-axis
alignments, we decided to test in one example if any remaining structure
coincide with other characteristic directions of the LSC. When dealing
with an individual cluster we fit an axi-symmetric LOS anisotropy model
involving the next quadratic term of the anisotropy expansion in $\mu$.
Such a procedure enables a better subtraction of the LOS effect, and thus
gives us a chance to look for any galaxy alignment non-coincident with
this axis. Thus, instead of the distribution (4.1) we use one involving
two first terms of the expansion of $F(\mu)$ in powers of
$\mu$\footnote{One should note, that the data for non-spiral galaxies in
the Virgo cluster does not allow for a fit with the model (4.1) providing
the LOS effect amplitude as large as the real one.}:
 
$$F(\mu) = {1 \over 1+A+B}+{2A \over 1+A+B}\, \mu + {3B \over 1+A+B}\,
\mu^2 \quad . \eqno(D.1) $$
 
\noindent
We fit this model to the data for our cluster with the largest number of
galaxies, the Virgo cluster (Cluster 11), in the following way. For  any
set of parameters $A$ and $B$ from the ranges $0 \le A \le 2A_{LSC}$ and
$-A_{LSC}  \le  B  \le  A_{LSC}$  we  derived  an  axi-symmetric   model
distribution for test galaxies placed in the positions of the real  ones
($A_{LSC}$ is the  amplitude  fitted  within  the  model  (4.1)  to  the
respective -- ALL, S, NS -- data  derived  in  section  4  for  all  LSC
galaxies). For any such galaxy distribution,  a  map  of  the  alignment
coefficient  $\Delta_{11}   /\sigma(\Delta_{11})$   was   provided   and
subtracted from the  map  for  the  real  cluster.  The  solutions  were
selected providing the smallest residuals on the map  (Fig.~5).  On  the
left side of the figure we presented the maps for real cluster while on
the left side we presented the residuals. Of course, our maps  generated
with the use of the model are subject to  statistical  fluctuations  and
the described procedure provides only a good approximation to  the  best
set ($A$, $B$).
 
Inspection of Fig.~5 shows that subtraction of  the  model  distribution
does not leave any significant residues for all galaxies ($A = 0.18$, $B
= 0.37$) and for spiral galaxies ($A = 0$, $B = 0.13$),  when  a  rather
weak 2-$\sigma$ alignment for NS  galaxies  ($A  =  1.6$,  $B  =  2.8$),
non-coincident with any of the tested directions was found. In this last
case the significance of the residual alignment is even smaller than the
cited value due to the large number of directions tested. Without a much
more thorough study one is unable to state firmly if any real  alignment
exists in this cluster.
 
\renewcommand{\thefigure}{\arabic{section}}
\setcounter{figure}{5}
\setcounter{section}{5}
 
\begin{figure*}
\vskip 11cm
\includegraphics{suraa.ps}
\caption{  Cluster  11:  In  the  left  panels  the  original  data  are
given  and  in  the  right  ones  the  residua  of   our   best   fitted
models are presented.}
\end{figure*}

\end{document}